\documentclass[twocolumn,showpacs,preprintnumbers,amsmath,amssymb]{revtex4}
\usepackage{dcolumn}
\usepackage{bm}
\usepackage{graphicx}
\usepackage{scrextend}

\begin{document}

\preprint{}

\title{Hole spin resonance and spin-orbit coupling \\ in a silicon metal-oxide-semiconductor field-effect transistor}
\author{K. Ono$^{1}$\footnote{E-mail address:
k-ono@riken.jp}\footnote{\label{note1}these authors contributed
equally to this work},
G. Giavaras$^{2}$\footref{note1}, 
T. Tanamoto$^{3}$, T. Ohguro$^{3}$, Xuedong Hu$^{2,4}$, and F.
Nori$^{2,5}$} \affiliation{$^1$Advanced device laboratory, RIKEN,
Wako-shi, Saitama 351-0198, Japan} \affiliation{$^2$CEMS, RIKEN,
Wako-shi, Saitama 351-0198, Japan} \affiliation{$^3$Corporate R\&D
Center, Toshiba Corporation, Kawasaki-shi, Kanagawa 212-8582,
Japan} \affiliation{$^4$Department of Physics, University at
Buffalo, SUNY, Buffalo, New York 14260-1500, USA}
\affiliation{$^5$Department of Physics, The University of
Michigan, Ann Arbor, MI 48109-1040, USA}

\date{\today}

%
\begin{abstract}
We study hole spin resonance in a p-channel silicon
metal-oxide-semiconductor field-effect transistor. In the
sub-threshold region, the measured source-drain current reveals a
double dot in the channel. The observed spin resonance spectra
agree with a model of strongly coupled two-spin states in
the presence of a spin-orbit-induced anti-crossing. Detailed
spectroscopy at the anti-crossing shows a suppressed spin
resonance signal due to spin-orbit-induced quantum state mixing.
This suppression is also observed for multi-photon spin
resonances. Our experimental observations agree with theoretical
calculations.
\end{abstract}


\maketitle


The silicon-based metal-oxide-semiconductor field-effect
transistor (MOSFET) is a key element of large-scale integrated
circuits that are at the core of modern technology. Looking into
the future, a universal fault-tolerant quantum computer also
requires a huge number of physical qubits, on the order of $10^8$
or more~\cite{QubitIntegration, SC Review}. As such, a qubit
integrated with the standard Si MOSFET architecture would be truly
attractive from the perspectives of scaling up and leveraging
existing technologies. One example of such a qubit is the spin of
an impulity/defect in the channel of a Si MOSFET. Indeed, spin
qubits defined in Si nano-devices are not only compatible with
current silicon technology, but are also known to be one of the
most quantum coherent among known qubit
designs~\cite{JelezkoPRL2004, MorelloPRB2009, MorelloNature2010,
KoehlNature2011, MortonReview2011, ZwanenburgReview2013,
MuhonenNN2014, VeldhorstNN2014, pratiNN2012, buluta, zalba16}.

Although there are many studies of impurities and defects in
Si~\cite{CompilationDefects}, single impurity/defect in the
channel of a Si MOSFET has only recently been studied
experimentally, by single-electron tunneling~\cite{Dot in MOS1,
Dot in MOS2, Dot in MOS3, Dot in MOS4, Dot in MOS Ono}. Spins of
such defects are difficult to characterize because of their
weakly-interacting nature. Controlling the spins of impurities in
a MOSFET, as well as in a gate-confined quantum dot, can be
achieved much more easily in a p-channel MOSFET than an n-channel.
The reason is that the larger spin-orbit interaction (SOI) of a
hole (-like) spin enables the spin resonance by an oscillatory
electric field, instead of a magnetic field, at microwave
frequencies under typical sub-Tesla static magnetic fields. Such
electrically-driven spin resonance (EDSR) has been demonstrated in
III-V devices~\cite{GsAs EDSR, III-V nanowire0, III-V nanowire,
III-V nanowire2}, as well as in Si~\cite{CMOS Si ESR,
kawakami14, takeda16}, while SOI effects in gate-confined Si
quantum dots have been investigated in the spin blockade
region~\cite{PMOS SB}. However, systematic investigations of
EDSR under the direct influence of SOI have not been performed in
Si, the material that provides an ideal stage for studying SOI due
to the minor presence of nuclear spins.

In this work we study sub-threshold transport and EDSR in a
short p-channel Si MOSFET, and quantitatively reveal the effects
of SOI and EDSR on lifting the spin blockade. Specifically, our
transport measurements demonstrate that there are two effective
dots in the channel, which allow us to identify a spin blockade
regime and explore spin resonance for two strongly-coupled
holes. The observed two-spin EDSR spectra, in particular the magnetic
field dependence of the resonances, and the associated state
mixing provide clear evidence of a SOI-induced anti-crossing with
a well-resolved spin-orbit gap. Spectroscopy at the anti-crossing
shows a suppressed EDSR signal because the involved states
are almost equally populated as a result of the maximum
SOI-induced state mixing. Our observations of spin blockade,
single- as well as multi-photon spin resonance, and
spin-orbit-induced state mixing are important steps toward the
precise control of spin qubits in Si MOSFETs.

\begin{figure}[t]
\begin{center}
\includegraphics[width=0.45 \textwidth, keepaspectratio]{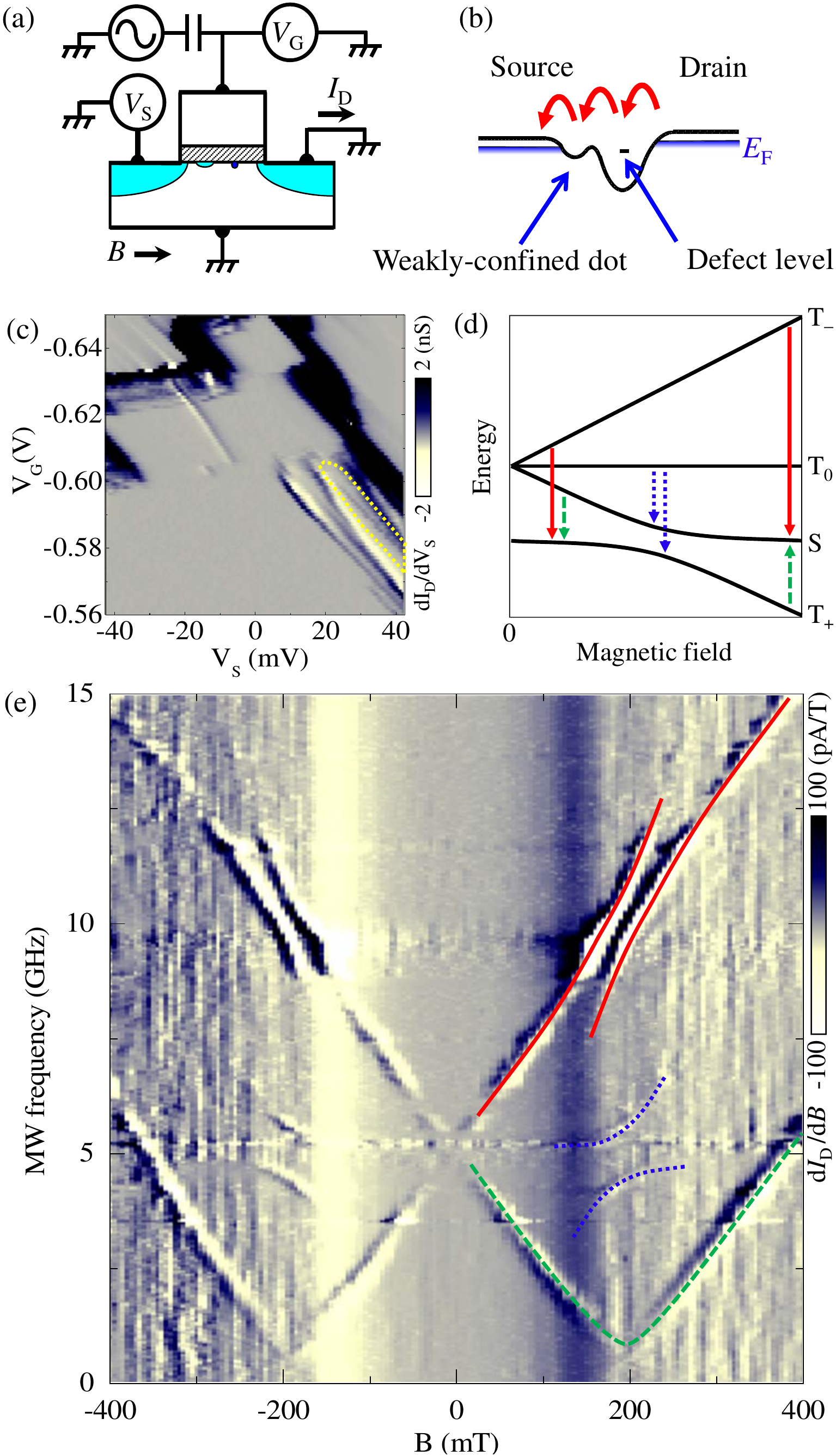}
\caption{(a) Schematic of the MOSFET device and measurement
set-up. (b) Potential landscape of quantum dots. (c)
Intensity plot of $dI_{D}/dV_{S}$ near the sub-threshold region.
The spin resonance is observed in the region enclosed by the
yellow dotted curve. The $dI_{D}/dV_{S}=0$ regions at the two
upper corners are artifacts of the current meter. (d) Schematic
energy diagram for two-hole states with a $T_{+}$--$S$
anti-crossing due to the spin-orbit interaction. The
microwave-induced transitions $T_-$--$S$ (red solid), $T_{0}$--$S$
(blue dotted), $T_{+}$--$S$ (green dashed) are indicated by
vertical arrows. (e) Intensity plot of $dI_{D}/dB$ measured at
$V_{S} = 25$ mV, $V_{G}= -0.597$ V. For $B>0$ the
high-current EDSR curves due to the transitions $T_-$--$S$ (red
solid), $T_{0}$--$S$ (blue dotted), $T_{+}$--$S$ (green dashed)
are indicated. Plotting $dI_{D}/dB$ suppresses resonances at
constant frequency due to photon-assisted tunneling enhanced by
cavity modes.}
\end{center}
\end{figure}

Our device is a p-channel MOSFET with a channel length of 135 nm
and width of 220 nm, as shown in Fig. 1(a). It has a silicon
oxynitride gate dielectric, and is fabricated with standard 0.13
$\mu$m CMOS technology. The measurements are performed in a
$^{4}$He pumped cryostat at a temperature of $T=1.6$ K. A magnetic
field is applied parallel to the MOS interface and the
source-drain current, and a microwave field is applied directly to
the gate electrode. Figure 1(c) shows the measured current in the
sub-threshold region. Specifically, we measure the source-drain
differential conductance as we vary the source-drain ($V_S$) and
gate ($V_G$) voltages. A Coulomb diamond with charging energy of
25 meV is observed centered around $V_{G}=-0.62$ V. The current in
this diamond is about three orders of magnitude smaller than the
on-state current of the MOSFET, which is a clear evidence of
Coulomb blockade. This has been observed in MOSFETs before and
attributed to sequential tunneling through a single dopant/defect
in the channel~\cite{Dot in MOS1, Dot in MOS2, Dot in MOS3, Dot in
MOS4, Dot in MOS Ono}.

An important feature of Fig.~1(c), however, is that the Coulomb
diamond around $V_{G}=-0.62$ V does not close all the way to $V_S
= 0$ at both its ends near $V_G = -0.60$ V and $V_G = -0.63$ V
(this is particularly clear near $V_G = -0.60$ V). This indicates
the presence of a larger dot that is detuned from and coupled in
series with a more tightly confined dot, so that sequential
tunneling through the double dot can only take place at finite
source-drain bias. The data in Fig.~1(c) indicates that the two
dots have a weak ($\sim 5$ meV) and a strong ($\sim 25$ meV)
confinement. The strongly-confined dot could be a Boron dopant in
the channel or a dangling bond defect at the silicon/oxynitride
interface, whereas the weakly-confined dot could arise from
potential fluctuations caused by remote impurities/defects. The
physical system can then be represented schematically as shown in
Fig.~1(b). Thermal cycles between 1.6 K and 300 K slightly shift
the gate voltage dependence, but the Coulomb diamond and the
microwave spectroscopy data remain the same after the cycles,
indicating the robustness of the double dot.

An interesting regime of double quantum dots is the spin blockade
regime, where spin symmetries are correlated with charge
configurations~\cite{SB}. In our double dot device, we have
evidence of spin blockade. Recall that in the spin
blockade~\cite{SB} transport is blocked if the two-spin state is
one of the triplet states, $T_{-}$, $T_{0}$, or $T_{+}$. Lifting
the spin blockade requires cotunneling and/or spin relaxation to
the singlet state $S$ that consists of $S_{11}$ and $S_{02}$
components~\cite{Petta_Science05}. Specifically, in the area
enclosed by the dotted curve in Fig.~1(c), the current is
suppressed outside the Coulomb blockade diamond, which indicates
that details of the electronic states, such as spin symmetry,
prevent electrons from sequential tunneling. Further evidence of
spin blockade is revealed when the suppression of conduction is
lifted by a microwave applied to the gate electrode, and
well-defined current peaks appear depending on both the external
magnetic field and the microwave frequency [Fig.~1(e)]. These
microwave-induced peaks define the high-current curves seen in
Fig.~1(e), and are due to spin excitations that lift the spin
blockade which was originally in place. No EDSR was observed on
the opposite side of the Coulomb diamond, for $V_{S} < 0$, mostly
because the tunneling is asymmetric for a MOSFET that is forward-
and reverse-biased.

The spectroscopic features of Fig.~1(e) can be qualitatively
explained by the low-energy spectrum of two-hole spin states in a
double dot [Fig.~1(d)], and also dove-tail nicely with the picture
of current suppression due to spin blockade. In our double dot
there is a singlet-triplet exchange splitting at zero magnetic
field due to mixing between the $S_{11}$ and $S_{02}$
singlets~\cite{Petta_Science05}. When a finite magnetic field is
applied, the triplet states Zeeman-split, with one of the
polarized triplets eventually crossing the singlet state. The SOI
couples the $T_{+}$ triplet with the $S_{02}$ singlet and makes
the crossing point into an anti-crossing. The magnitude of the
anti-crossing gap is determined by the SOI matrix element between
the $T_{+}$ and the $S_{02}$, and in our device it is about 1 GHz.
The two eigenstates near the anti-crossing are mostly mixtures of
$S_{11}$ and $S_{02}$ singlets together with the $T_+$ triplet.
The field at which the anti-crossing occurs, i.e., $\pm 200$ mT in
Fig.~1(e), is determined by the zero-field exchange splitting and
the $g$-factors in the two dots.

The high-current curves in Fig.~1(e) can now be attributed to
microwave-induced transitions between the mixed singlet-triplet
states as indicated by the arrows in Fig.~1(d). Microwave-induced
transitions among the triplet states ($T_{\pm}$ to $T_0$, i.e. the
normal EDSR transitions) do not lift the spin blockade, thus
cannot be observed in our transport experiment. SOI does not
couple $T_0$ and $S$ states, thus we do not observe a horizontal
current curve in Fig.~1(e), except near the anti-crossing, where
the $T_0$ to $T_+$ transition is allowed and the spin blockade can
be lifted because of the $T_{+}$--$S$ mixing. Similar EDSR curves
have also been observed in III-V nanowire double dots~\cite{III-V
nanowire}. Notice that, in Fig.~1(e) the background current
increases at $\pm 200$ mT, independent of the microwave frequency,
giving a clear vertical contrast at these fields. This increase is
consistent with the enhanced scattering rate due to the
SOI-induced $T_{\pm}$--$S$ mixing.

The EDSR spectra up to 40 GHz indicate that the $g$-factor
difference between the two dots is small compared with the
zero-field singlet-triplet splitting of about 5 GHz~\cite{Suppl}.
Assuming the same $g$-factors, the slope of the current curves in
Fig.~1(d) gives a $g$-factor of 1.80. This is much larger than the
value 1.1 observed for Boron dopants in bulk Si~\cite{Boron
g-factor}, while smaller than the value 2.0 of the dangling bond
defect centers at the silicon-oxynitride interface~\cite{K center
g-factor}. We generally expect shallower defects to be more
affected by the spin-orbit nature of the valence band, and their
$g$-factors should be smaller than the value of deep dangling bond
defects. EDSR spectra as in Fig.~1(e) can be observed throughout
the spin blockade area enclosed by the dotted curve in Fig.~1(c).
The $g$-factor does not change significantly in this area, while
the exchange energy can change by a factor of 2 depending on
$V_{G}$. The typical linewidth of EDSR is 0.18 GHz, probably
limited by the electrical charge noise due to the strong SOI in
our device.We expect only a minor contribution of the nuclear
spins to the EDSR linewidth due to the small content (4\%) of
29-Si, and the p-orbital nature of holes.

\begin{figure}[t]
\begin{center}
\includegraphics[width=0.48 \textwidth, keepaspectratio]{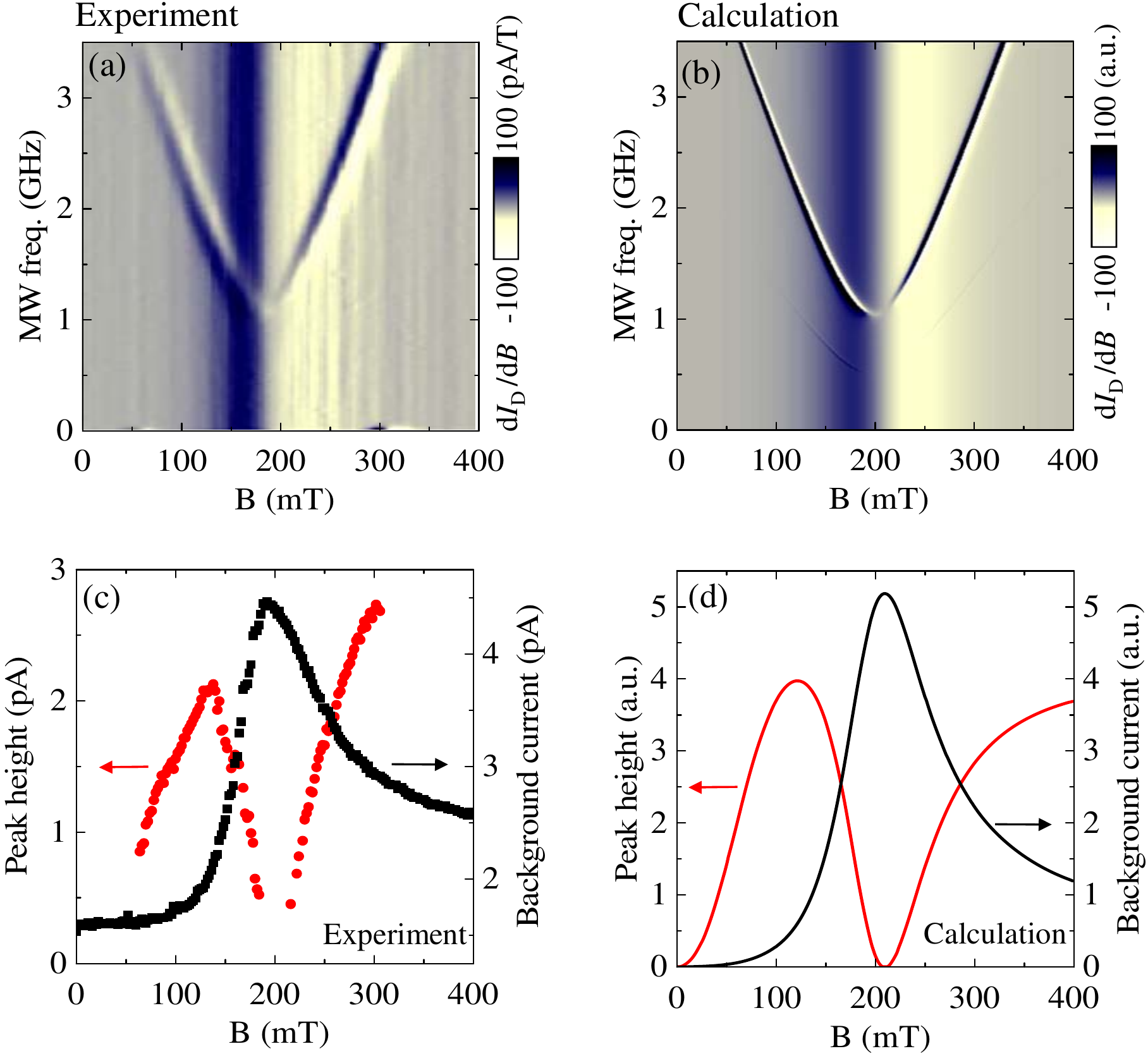}
\caption{(a) Measured and (b) calculated spin resonance spectra
near the $T_{+}$--$S$ anti-crossing point for weak microwave
driving [$-40$ dBm at the output of the microwave source for
(a), and microwave amplitude $A=30$ $\mu$eV for (b)]. Measured (c)
and calculated (d) peak height (bright line, left axis), and
background current without microwave (dark line, right axis).}
\end{center}
\end{figure}

It is emphasized that our experiment is performed at
temperature of 1.6 K, which is over an order of magnitude higher
than the usual temperatures of 0.1 K reported in previous works
[20-26]. Performing the experiment at this high temperature is
achieved thanks to the large orbital quantization energy of our
dots. This gives also tolerance against unwanted
photon-assisted tunneling or pumping current under strong
driving.

For a more precise understanding of our observations, we focus on
the $T_{+}$--$S$ transition near the anti-crossing point. This
anti-crossing has never been observed before; neither in Si nor in
III-V quantum dots. Figure~2(a) shows the leakage current
($dI_{D}/dB$) as a function of the microwave frequency and the
magnetic field. The physics here can be well explained by a
two-level model described in the Supplement~\cite{Suppl}. To
summarize briefly, we incorporate the microwave driving by
assuming that an electric field of amplitude $A$ and frequency
$f=\omega/2\pi$ modulates the on-site energy $\varepsilon_2$ of
dot 2 periodically, namely,
$\varepsilon_2\rightarrow\varepsilon_2+A\cos(\omega t)$. In other
words, the transitions we study are purely electrically driven.
The model considers the two energy levels $E_{1}$ and $E_{2}$
which anti-cross. The corresponding eigenstates are $|u_i\rangle =
a_i |S_{11} \rangle + b_i |T_{+}\rangle + c_i |S_{02}\rangle +
d_{i}|T_{-}\rangle$, $i=1$, $2$. The double dot parameters for
$A=0$ are chosen so that the levels anti-cross at about $200$ mT,
with a spin-orbit gap of about $1$ GHz. The coefficients $a_i$,
$b_i$, $c_i$, and $d_{i}$ are obtained by diagonalizing the double
dot Hamiltonian in the absence of the microwave. When the
microwave is turned on, we perform a unitary transformation into a
rotating frame~\cite{Suppl}, and within a rotating wave
approximation we obtain an approximate time-independent
Hamiltonian for the single-photon spin resonance
\begin{equation}
h_{\mathrm{DQD}}=
\left(%
\begin{array}{cc}
  E_{1}+\hbar\omega/2 & q  \\
  q & E_{2}-\hbar\omega/2 \\
\end{array}%
\right),
\end{equation}
with
\begin{equation}
q =  \hbar\omega\frac{c_1 c_2 }{(c^{2}_{1}-c^{2}_{2})}
J_{1}\!\left(\frac{A(c^{2}_{1}-c^{2}_{2})}{\hbar\omega}\right),
\end{equation}
where $J_1$ is the $1$st order Bessel function of the first
kind~\cite{Suppl}. We then calculate the current with a density
matrix approach~\cite{Suppl}.

The theoretical results from this two-level model, shown in
Fig.~2(b), are in good qualitatively agreement with the
experimental observations in Fig.~2(a). There are two important
features common to both figures, one being the broad peak in the
background current ($A=0$) centered at about $200$ mT independent
of the microwave frequency. This peak is the result of the
SOI-induced singlet-triplet mixing. It has an asymmetric form
\cite{Giavaras}, unlike the usual symmetric lineshape in a
two-level system. The other common feature of Figs.~2(a, b) is the
high-current curve due to the microwave-induced $T_{+}$--$S$
transition. The shape of this curve is hyperbolic, which arises
from the normal anti-crossing of two straight lines. The two-level
model we adopt here gives us a good qualitative description of the
experimental observations. We do not attempt to achieve
quantitative agreement because of the missing information with
regard to the device, such as the exact interdot tunnel coupling
and the microscopic spin-orbit coupling mechanism. For example,
differences in the EDSR linewidths between experiment and theory
are most probably due to different co-tunneling rates that limit
the lifetime of spin states in the dots, as well as additional
decoherence sources that are not accounted for in the model.

Experimental data in Fig.~2(c) demonstrate that near the
anti-crossing at $200$ mT the background current reaches a
maximum, while the EDSR-induced current has a minimum. This
minimum occurs even though the transition rate between the two
levels due to the microwave field is the highest because of the
maximized singlet-triplet mixing. This interesting feature can be
understood within the two-level model. Recall that the leakage
current in the spin blockade is due to mixing of the triplet with
the singlet state. The microwave field indeed tends to equalize
the occupations of the two levels, but near the anti-crossing the
SOI already generates the maximum possible singlet-triplet mixing,
so that transport of the electrons occupying the $T_+$ triplet
state is no longer blocked. This leads to a maximum in the leakage
current, and transitions between the two states due to the
microwave field cannot increase the current further. Thus the
effect of the microwave is almost completely suppressed. This
situation is similar to the well-known saturation of absorption
under strong driving in spin resonance experiments~\cite{ESR},
where the microwave equalizes the populations of the two levels
and eventually leads to a decrease in the resonance signal.
Figure~2(d) shows the calculated single-photon EDSR-induced
current peak height as well as the background current, which are
in nice qualitative agreement with the experimental observations
in Fig.~2(c). Notice that while at the $T_{+}$--$S$ anti-crossing
the microwave-induced $T_{+}$--$S$ transition does not lead to
further increase in current, the $T_{0}$--$S$ and $T_{-}$--$S$
transitions do lead to a current increase because they lift the
spin blockade for electrons occupying the $T_-$ and $T_0$ states.
In Fig.~1(c), the microwave-induced current increase is visible
even at $\pm 200$ mT.

\begin{figure}[t]
\begin{center}
\includegraphics[width=0.45 \textwidth, keepaspectratio]{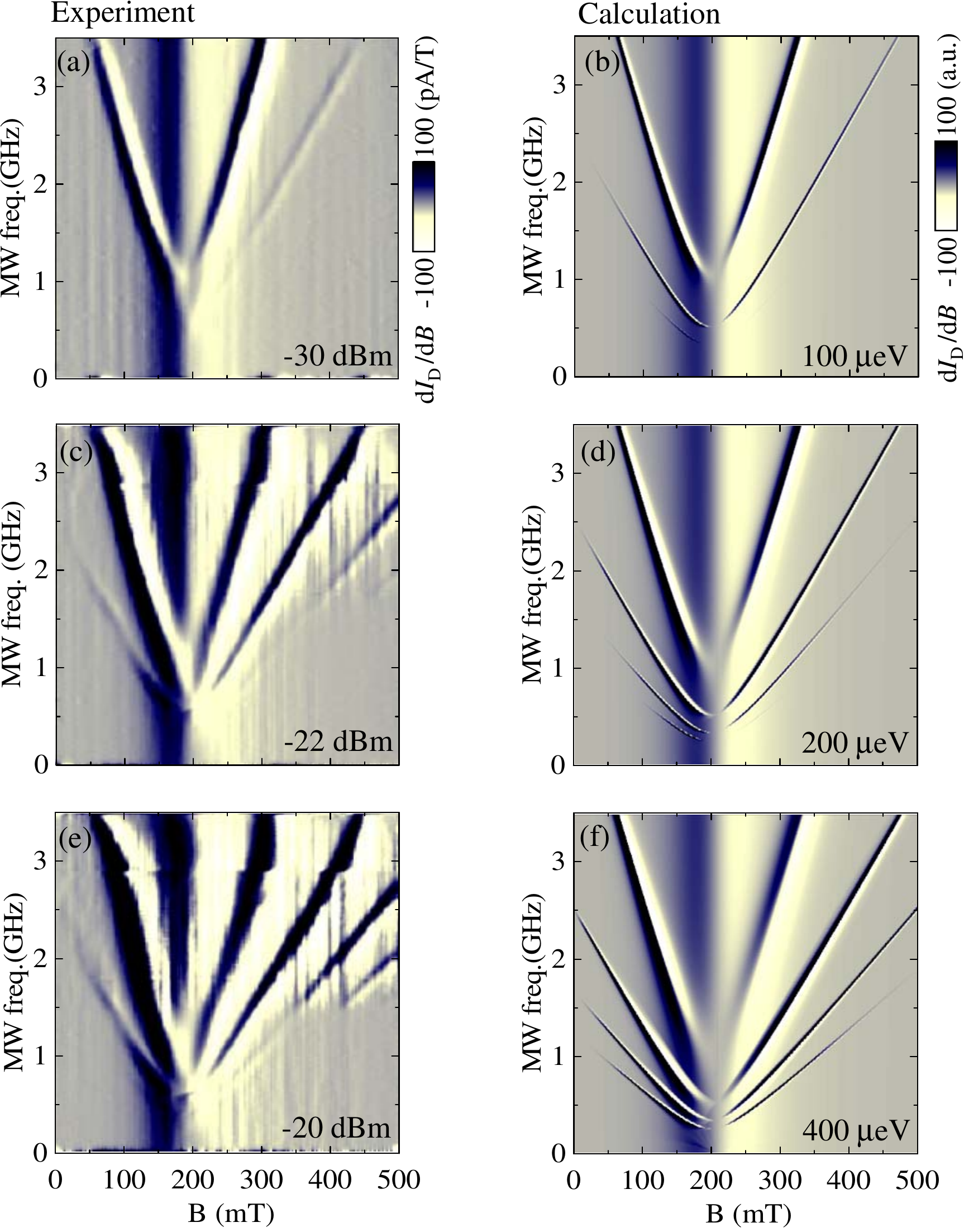}
\caption{Measured spin resonance spectra near the $T_{+}$--$S$
anti-crossing at higher microwave powers (a) $-30$ dBm, (c) $-22$
dBm and (e) $-20$ dBm at the output of the microwave source
respectively. Results of calculation for microwave amplitude (b)
$A=100$ $\mu$eV, (d) $A=200$ $\mu$eV, and (f) $A=400$ $\mu$eV.}
\end{center}
\end{figure}

In Fig.~2 the EDSR-induced current peaks also diminish for $B
\rightarrow 0$ [the feature is also apparent in Fig.~1(e)]. Within
the two-level model, when $B\rightarrow0$ the state $|u_2\rangle$
becomes more exclusively the polarized triplet state, $|u_2\rangle
\approx |T_{+}\rangle$, so that the coupling term $q\rightarrow 0$
because $c_2 \rightarrow 0$. Thus, the microwave field becomes
less efficient in inducing direct transitions from any of the
triplet states to the singlet and the current peaks start to
diminish for $B\rightarrow0$. A cautionary note here, however, is
that the two-level model becomes increasingly inaccurate as $B
\rightarrow 0$, because in this limit the triplets become quasi
degenerate. In the Supplement~\cite{Suppl} we discuss a more
accurate calculation based on a Floquet master equation, which
confirms the trends observed in Fig.~2.

Multi-photon EDSR has been observed before in double dots at
strong microwave driving~\cite{2 Photon coherent ESR, Multi-photon
ESR}, away from the $T_+$--$S$ anti-crossing. As shown in
Figs.~3(a, c, e), when we increase the microwave power in our
device we can generate additional current peaks. These peaks
correspond to $n=2$, 3, or more photons inducing transitions
between the two levels that anti-cross. The resulting multi-photon
high-current curves are extrapolated to the $1/n$ of the
spin-orbit gap at $200$ mT. The multi-photon peaks can be
reproduced with the two-level model discussed above when we use
the appropriate $n$-photon Hamiltonian~\cite{Suppl}. The
theoretical results in Figs.~3(b, d, f) are in good qualitatively
agreement with the experiment. Increasing the microwave amplitude
$A$ gives rise to extra current peaks, in addition to the primary
single-photon one, corresponding to the successive $n$-photon
resonance $n\hbar\omega = E_{2}-E_{1}$. Here results up to
four-photon transitions are shown. As derived in the
Supplement~\cite{Suppl} the Hamiltonian describing the $n$-photon
transition depends on $n$. Therefore, in Figs.~3(b, d, f) we
consider $1\le n \le 4$, and for each frequency we plot the
corresponding maximum increase in the background current that
comes from a specific $n$. This way produces the correct behaviour
near the $n$-photon peak.

In summary, we studied a p-channel Si MOSFET and identified a spin
blockade regime in a double dot system formed by a pair of
defects/impurities in the channel. We experimentally observed
electrically-driven two-spin resonance and found that the spin-orbit
interaction suppresses the spin resonance signal near the
anti-crossing point for both single- and multi-photon resonances.
Our work shows that impurities/defects in commercial-quality Si
MOSFET can be addressed straightforwardly, and they provide a
useful window into the electronic spectrum and quantum coherent
dynamics. This revelation is particularly appealing when we
consider the great practical advantages that silicon industry
could provide to fabricating quantum coherent devices.

We thank M. Kawamura, K. Ishibashi, K. Itoh, S. Kohler, and S.
Shevchenko for discussions. This work was supported by JSPS
KAKENHI Grant No. 15H04000. This work was partially supported by
the RIKEN iTHES Project, the MURI Center for Dynamic
Magneto-Optics via the AFOSR Award No. FA9550-14-1-0040, the Japan
Society for the Promotion of Science (KAKENHI), the IMPACT program
of JST, CREST, US ARO, and a grant from the John Templeton
Foundation.

\appendix

\section{Double quantum dot: Coulomb diamond and current}

Most of the features of the open Coulomb diamond structure in
Fig.~1(c) in the main article can be well reproduced by a simple
calculation based on the constant charging-energy model. If
$N_{i}$ ($i=1$, $2$) is the number of holes on dot $i$, then the
energy of dot 1 is $E_{1}(N_{1}, N_{2}) = E_{C1} N_{1} +
E_{C12}N_{2} - C_{1}V_{G} - D_{1}V_{S} + E_{\rm off}$, and the
energy of dot 2 is $E_{2}(N_{1}, N_{2}) = E_{C12} N_{1}+E_{C2}
N_{2}-C_{2}V_{G}-D_{2}V_{S}$. Here, $E_{Ci}$ ($i=1$, 2) and
$E_{C12}$ denote an on-site and an inter-dot charging energy
respectively. Also, $C_i$ and $D_i$ are the lever arms of $V_G$
and $V_S$, while $E_{\rm off}$ is the energy offset between the
dots. The Coulomb blockade is lifted for $eV_{S} > E_{1}(N_{1}+1,
N_{2}) > E_{1}(N_{1}, N_{2}+1) > 0 (= eV_{D})$. Figure~S1 shows
the typical Coulomb diamond structure for a double quantum dot
when one of the dots has large charging energy, and the other dot
has small charging energy.
\begin{figure}[t]
\begin{center}
\includegraphics[width=0.45 \textwidth, keepaspectratio]{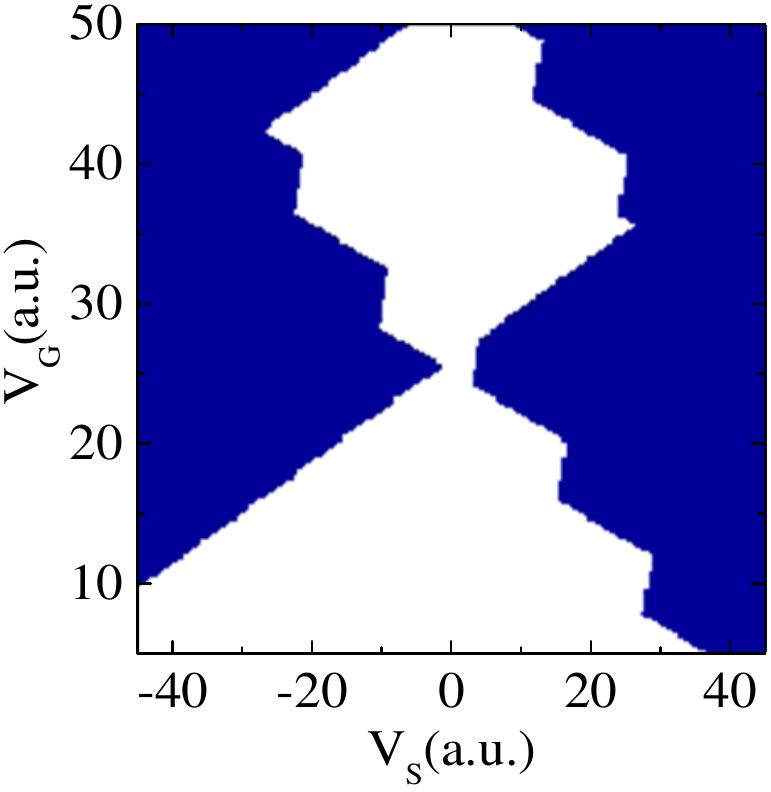}
\caption{Coulomb diamond structure for a double quantum dot
calculated by the constant charging model. The white region
corresponds to the Coulomb blockade region. The parameters (a.u.)
are: $E_{C1} = 5$, $E_{C2} = 25$, $E_{C12} = 0.2$, $E_{\rm off} =
-0.25$, $C_{1} = 1.1$, $C_{2} = 1.0$, $D_{1} = 0.33$, $D_{2} =
0.66$.}
\end{center}
\end{figure}

In Fig.~1(c) in the main article, a region where spin blockade
occurs was identified. The transport cycle in the spin blockade
regime is shown schematically in Fig.~S2. As explained in the main
article the spin-orbit interaction and the microwave field can
lift the spin blockade by inducing singlet-triplet transitions.
\begin{figure}[h]
\begin{center}
\includegraphics[width=0.45 \textwidth, keepaspectratio]{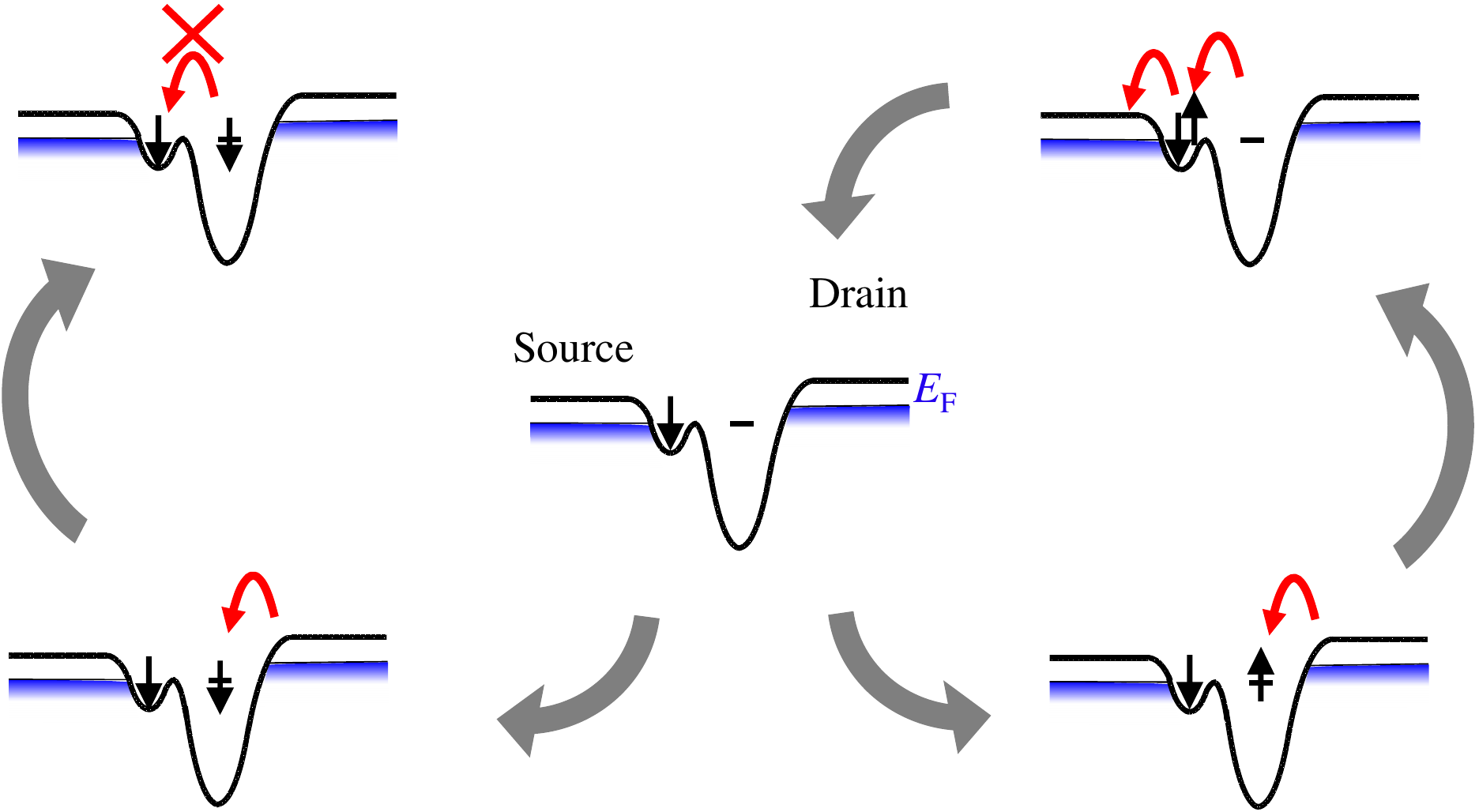}
\caption{Schematic representation of the transport cycle in the
spin blockade regime. If a state in the bias window has no double
occupation on the left dot the current is blocked. The spin-orbit
interaction and the microwave field can lift the spin blockade by
inducing singlet-triplet transitions.}
\end{center}
\end{figure}
As a result a measurable leakage current flows through the double
dot. Figure~S3 shows the intensity plot of the leakage current
$I_{D}$ for the same scale of magnetic field $B$ and MW frequency
$f$ as that in Fig.~1(e) in the main article (where $dI_{D}/dB$
was presented). The high-current curves are due to
microwave-induced transitions between the mixed singlet-triplet
states. The series of resonances at constant frequency are due to
photon-assisted tunneling enhanced by cavity modes.
\begin{figure}[t]
\begin{center}
\includegraphics[width=0.45 \textwidth, keepaspectratio]{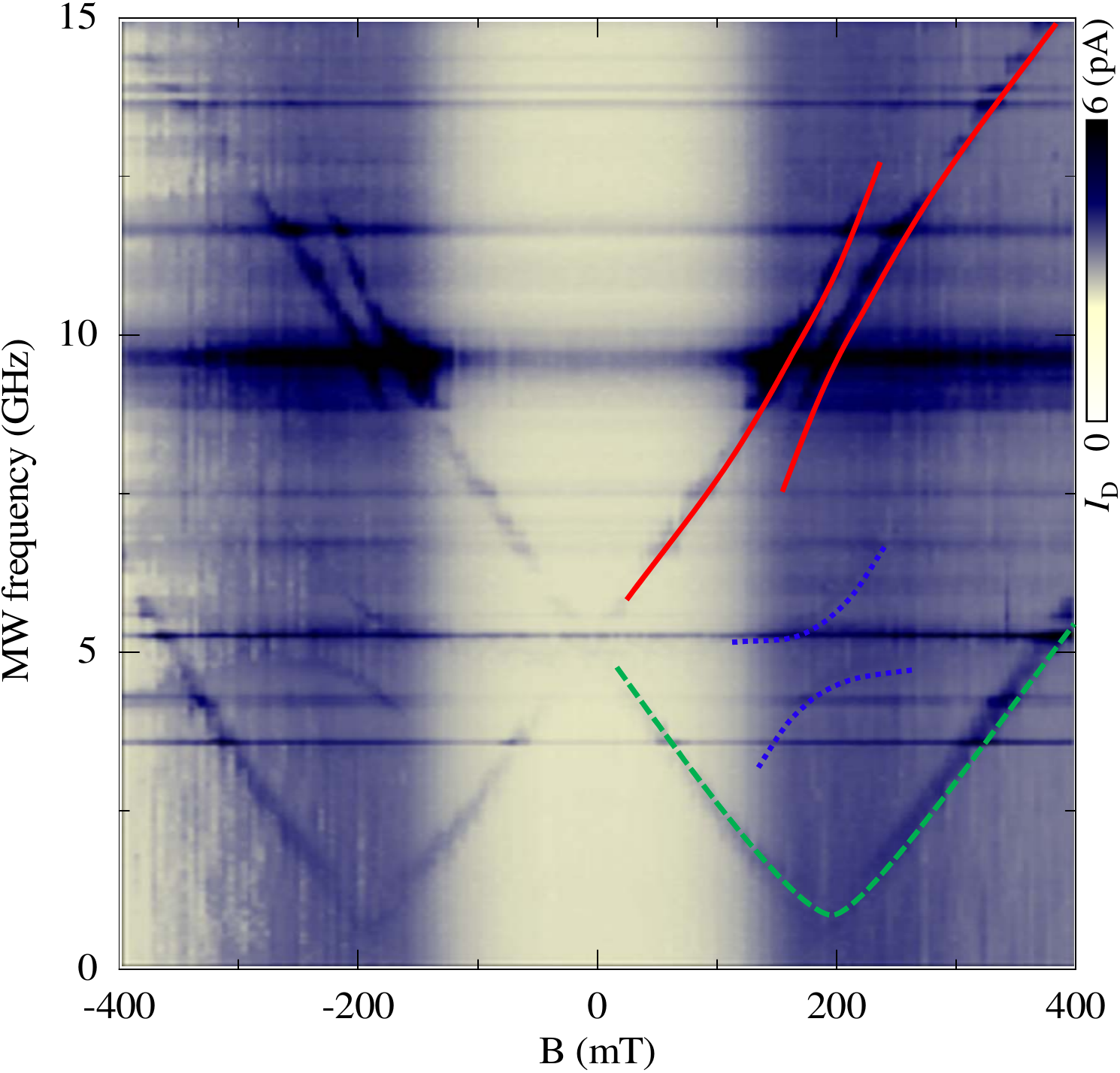}
\caption{Intensity plot of leakage current $I_{D}$. For $B>0$ the
high-current EDSR curves due to the transitions $T_-$--$S$ (red
solid), $T_{0}$--$S$ (blue dotted), $T_{+}$--$S$ (green dashed)
are indicated.}
\end{center}
\end{figure}

\section{Spin resonance for large magnetic field}

In the main article we presented EDSR spectra near the
$T_{+}$--$S$ anti-crossing point. Here we show additional spectra
for a microwave frequency up to 40 GHz and magnetic field up to
1.7 T. In Fig.~S4 three nearly-straight lines are visible. As
explained in the main article, two of these lines map-out the
transitions between the states $T_{\pm}$ and $S$. The lower line
corresponds to the 2-photon $T_{+}$--$S$ transition. For a double
quantum dot with large difference in the $g$-factors, the lines
$T_{\pm}$--$S$ are not parallel at high magnetic fields.
Investigation of the data shown in Fig.~S4 demonstrates that in
our system these lines are parallel within at least 2\% accuracy,
indicating that the $g$-factor difference in the two dots is small
enough compared with the zero-field singlet-triplet splitting of
about 5 GHz.

\begin{figure}[t]
\begin{center}
\includegraphics[width=0.45 \textwidth, keepaspectratio]{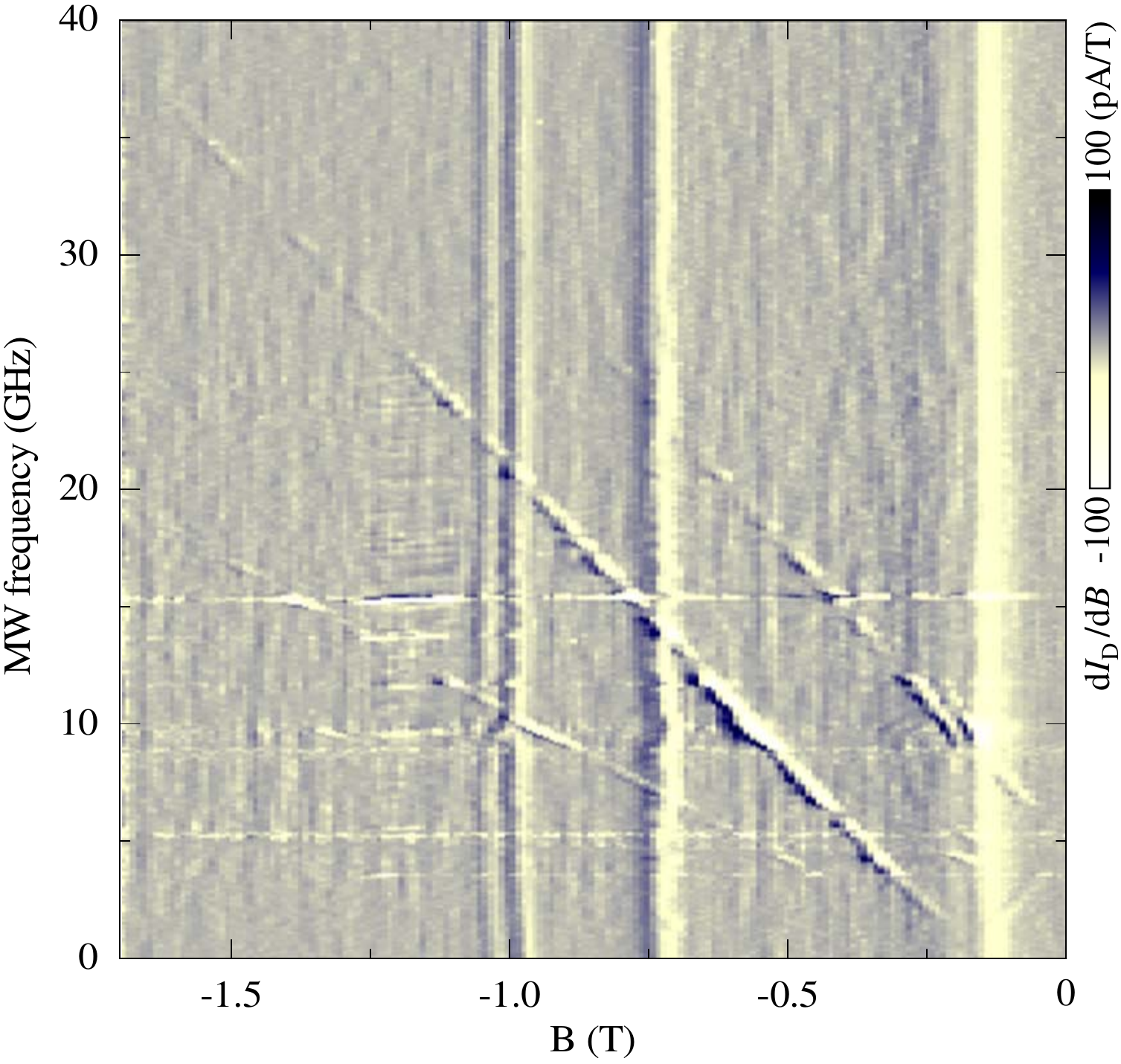}
\caption{Intensity plot of $dI_{D}/dB$ at high microwave frequency
and magnetic field.}
\end{center}
\end{figure}

\section{On the microwave attenuation and nonlinearity}

In this section we present some details about the microwave field.
If we assume a 50 Ohm impedance for our transmission line, then
the MW power (in dBm) used in the experiment [Fig.~2(a) and
Fig.~3(a,
c, e)], and the corresponding MW amplitude (in mV) are:\\
-40 dBm  2.2 mV,\\
-30 dBm  7.1 mV,\\
-22 dBm 17.8 mV,\\
-20 dBm 22.4 mV.\\
For the calculations [Fig.~2(b) and Fig.~3(b, d, f)] we have the
following correspondence:\\
-87 dBm 0.01 mV,\\
-67 dBm 0.1 mV,\\
-61 dBm 0.2 mV,\\
-55 dBm 0.4 mV.\\

The experimental and theoretical numbers are quite different,
suggesting a very large attenuation and nonlinearity. An estimated
attenuation of our rigid coaxial cable with a 2 m length is only 3
dB at 5 GHz. This very large attenuation and nonlinearity may be
due to the bare wiring of about 1 cm length between the end of the
coax and our device, as well as the nonlinearity of the
capacitance. Our MOSFET device is set around the subthreshold
regime, thus the capacitance will be affected by the voltages
$V_{S}$, $V_{G}$ as well as the MW power, if a strong power starts
to cause photon-assisted tunneling or charge pumping current.

\section{System Hamiltonian}

In this section we describe the Hamiltonian of the physical
system. We consider a double quantum dot (DQD) coupled to metallic
leads. The total Hamiltonian of the system is
\begin{equation}
H=H_{\mathrm{DQD}}+H_{\mathrm{L}}+H_{\mathrm{T}},
\end{equation}
where $H_{\mathrm{DQD}}$ is the DQD Hamiltonian, $H_{\mathrm{L}}$
is the Hamiltonian of the leads, and $H_{\mathrm{T}}$ is the
interaction Hamiltonian between the DQD and the leads.
Specifically, the DQD Hamiltonian is
\begin{eqnarray}
H_{\mathrm{DQD}}& = & \sum^{2}_{i=1}\left(\varepsilon_{i}n_{i}+U_{i}n_{i\uparrow}n_{i\downarrow}
-\frac{1}{2}g_{i}\mu_{B}B(n_{i\uparrow}-n_{i\downarrow})\right) \nonumber \\
& & + H_{\mathrm{c}} + H_{\mathrm{so}}, \label{ddot}
\end{eqnarray}
where $n_{i}$ is the number operator
$n_{i}=\sum_{\sigma}n_{i\sigma}=c_{i\uparrow}^{\dagger}c_{i\uparrow}+c_{i\downarrow}^{\dagger}c_{i\downarrow}$,
and the operator $c_{i\sigma}^{\dagger}$ ($c_{i\sigma}$) creates
(destroys) a hole on dot $i=1$, 2, with spin
$\sigma=\{\uparrow,\downarrow\}$ and orbital energy
$\varepsilon_{i}$. We assume a single-band description and
consider the holes to have spin 1/2. In this case the two-hole
Hilbert space is spanned by the singlet and triplet states
$|T_{\pm}\rangle$, $|T_{0}\rangle$, $|S_{11}\rangle$,
$|S_{20}\rangle$, $|S_{02}\rangle$, where $|S_{km}\rangle$ is a
singlet state with $k$ ($m$) holes on dot 1 (dot 2).

The orbital energies of the two dots are
\begin{equation}
\varepsilon_{1}= \frac{\delta}{2}, \qquad \varepsilon_{2}=  -U_{2}
- \frac{\delta}{2}+A\cos(\omega t),
\end{equation}
where $\delta$ denotes the energy detuning. The external electric
field has amplitude $A$ and cyclic frequency $\omega=2\pi f$, and
when $A\ne0$ the DQD Hamiltonian is time dependent
$H_{\mathrm{DQD}}=H_{\mathrm{DQD}}(t)$. This configuration of the
orbital energies results in a localised spin in dot 2 during the
transport cycle in the spin blockade regime.

Each dot has a charging energy $U_{i}$, and $g$-factor $g_{i}$
which leads to a Zeeman splitting $g_{i} \mu_{\mathrm{B}}B$ due to
the external magnetic field $B$. The inter-dot tunnel coupling
with strength $t_{\mathrm{c}}$ is modelled by the Hamiltonian
\begin{equation}
H_{\mathrm{c}}=-t_{\mathrm{c}} \sum_{\sigma}
c^{\dagger}_{1\sigma}c_{2\sigma}+\mathrm{H.c.},
\end{equation}
and the non spin-conserving inter-dot tunnel coupling due to the
spin-orbit interaction (SOI) is modelled by the
Hamiltonian~\cite{stehlik, Giavaras}
\begin{equation}\label{hso}
H_{\mathrm{so}}=-t_{\mathrm{so}}
\sum_{\sigma\sigma'}c^{\dagger}_{1\sigma}
(i\sigma^{y})_{\sigma\sigma'}  c_{2\sigma'}+\mathrm{H.c.}.
\end{equation}
This simplified Hamiltonian couples $|S_{02}\rangle$
($|S_{20}\rangle$) to $|T_{\pm}\rangle$ states, thus for example
the lowest singlet-triplet levels anti-cross and the induced gap
is proportional to the SOI tunnel coupling $t_{\mathrm{so}}$. For
a fixed $t_{\mathrm{so}}$ the anti-crossing gap is sensitive to
the detuning $\delta$ because this controls the amplitude of the
$|S_{02}\rangle$ component in the quantum states. A rigorous
derivation of a microscopic SOI Hamiltonian~\cite{golovach} should
consider the detailed geometry of the quantum dot system which in
the present device is unknown. Nevertheless, Eq.~(\ref{hso})
allows us to reproduce the basic experimental features.

The DQD is tunnel-coupled to left and right leads, which consist
of non-interacting holes. These holes are described by the
Hamiltonian
\begin{equation}
H_{\mathrm{L}}=\sum_{\ell k\sigma}\epsilon_{\ell k}d_{\ell
k\sigma}^{\dagger}d_{\ell k\sigma},
\end{equation}
where the operator $d_{\ell k\sigma}^{\dagger}$ ($d_{\ell
k\sigma}$) creates (destroys) a hole in lead $\ell=\{L, R\}$ with
momentum $k$, spin $\sigma$, and energy $\epsilon_{\ell k}$. The
interaction Hamiltonian between the DQD and the two leads is
\begin{equation}
H_{\mathrm{T}}=\sum_{k\sigma}(t_{L}c_{1\sigma}^{\dagger}d_{L
k\sigma} + t_{R}c_{2\sigma}^{\dagger}d_{R k\sigma}) +\text{H.c.},
\end{equation}
with $t_{L}$ $(t_{R})$ being the tunnel coupling between dot 1 (2)
and the left (right) lead, which is assumed to be energy
independent, and we also consider $t_{L}=t_{R}$.

\section{Two-level model}

\subsection{Two-level Hamiltonian}

In the main article an effective two-level Hamiltonian was used to
explore the microwave-induced peaks. Here we give some details
about the derivation of this Hamiltonian. First we diagonalize the
time-independent part of the DQD Hamiltonian $H_{\mathrm{DQD}}$.
The derived eigenenergies are denoted by $E_{i}$ and the
corresponding eigenstates are written in the general form
\begin{equation}
|u_i\rangle = a_i|S_{11}\rangle + b_{i}|T_{+}\rangle +
c_i|S_{02}\rangle + d_{i}|T_{-}\rangle + e_{i}|T_{0}\rangle.
\end{equation}
For only one state the coefficient $e_{i}\ne0$ and specifically
$e_{i}=1$, and for simplicity we neglect the component
$|S_{20}\rangle$, but this is taken into account in the numerical
computations. Then we write the total DQD Hamiltonian
$H_{\mathrm{DQD}}$ in the energy basis $|u_i\rangle$. To look for
an analytical treatment, we assume that the two eigenstates
$|u_{1}\rangle$, $|u_{2}\rangle$, which form the anti-crossing
point, can approximate well the dynamics of the system and thus we
ignore all the other eigenstates. These arguments lead to the
following approximate DQD Hamiltonian
\begin{equation}\label{2by2}
h^{'}_{\mathrm{DQD}}=
\left(%
\begin{array}{cc}
  E_{1} & 0 \\
  0 & E_{2} \\
\end{array}%
\right) + A\cos(\omega t)\left(%
\begin{array}{cc}
  1+c^2_{1} & c_{1}c_{2} \\
  c_{1}c_{2} & 1+c^2_{2} \\
\end{array}%
\right),
\end{equation}
where $E_{1}$, $E_{2}$ are the two energy levels which anti-cross.
Then to remove the time dependence from the diagonal elements of
$h^{'}_{\mathrm{DQD}}$, we perform a transformation to derive the
transformed Hamiltonian~\cite{grifoni}
\begin{equation}
h_{\mathrm{DQD}}=U^{\dagger}(t)h^{'}_{\mathrm{DQD}}U(t)-i\hbar
U^{\dagger}(t)\frac{dU(t)}{dt},
\end{equation}
with the operator
\begin{equation}
U(t)=\left(%
\begin{array}{cc}
  e^{i\phi_1(t)} & 0 \\
  0 & e^{i\phi_2(t)} \\
\end{array}%
\right),
\end{equation}
and the phases
\begin{equation}
\phi_{1,2}(t) = - \frac{(1+c^{2}_{1,2})A}{\hbar\omega} \sin(\omega
t) \pm \frac{n \omega t}{2}.
\end{equation}
The transformed Hamiltonian is
\begin{equation}
h_{\mathrm{DQD}}=
\left(%
\begin{array}{cc}
  E_{1}+n\hbar\omega/2 & q \\
  q^{*} & E_{2}-n\hbar\omega/2 \\
\end{array}%
\right),
\end{equation}
with the off-diagonal coupling element being
\begin{eqnarray}
q & = & \frac{c_1 c_2 A }{2} \left[\exp(+i \omega t) + \exp(-i\omega
t)\right] \nonumber \\
& & \times \exp(-i n \omega t) \exp\left( i
\frac{\Lambda}{\hbar\omega}\sin(\omega t) \right),
\end{eqnarray}
and the parameter $\Lambda= A (c^2_{1} - c^{2}_{2})$. To simplify
this expression we use the formula
\begin{equation}
\exp[ix\sin(\omega t)] =\sum_{m} \exp(i m \omega t)
J_{m}\left(x\right),
\end{equation}
where $J_{m}$ is the $m$th order Bessel function of the first
kind. Then the coupling term is
\begin{eqnarray}
q & = & \frac{c_1 c_2 A}{2}\sum_{m} \exp[i (m - n + 1) \omega t]
J_{m}\left(\frac{\Lambda}{\hbar\omega}\right) \nonumber \\
& & + \frac{c_1 c_2
A}{2}\sum_{m} \exp[i (m - n - 1) \omega t]
J_{m}\left(\frac{\Lambda}{\hbar\omega}\right). \label{qq1}
\end{eqnarray}
In the  context of a `rotating wave approximation', we assume that
in the long-time limit, when the system has reached the steady
state, the non-oscillatory terms can approximate well the
dynamics. Thus, the off-diagonal element becomes time-independent
\begin{equation}\label{qq2}
q \approx  \frac{c_1 c_2 A}{2}
J_{n-1}\left(\frac{\Lambda}{\hbar\omega}\right) + \frac{c_1 c_2
A}{2}J_{n+1}\left(\frac{\Lambda}{\hbar\omega}\right).
\end{equation}
Using the property $xJ_{n-1}(x)+xJ_{n+1}(x)=2nJ_{n}(x)$ and
substituting $\Lambda= A (c^2_{1} - c^{2}_{2})$, we arrive at the
off-diagonal coupling element
\begin{equation}
q = n \hbar \omega\frac{c_1 c_2}{c^2_{1} - c^{2}_{2}}
J_{n}\left(\frac{A (c^2_{1} - c^{2}_{2})}{\hbar\omega}\right),
\quad n=1, 2, ...
\end{equation}
We use the effective DQD Hamiltonian $h_{\mathrm{DQD}}$ to study
the $n$-photon resonance that satisfies the condition
$n\hbar\omega = E_{2}-E_{1}$. For $n=1$, the Hamiltonian
$h_{\mathrm{DQD}}$ coincides with the Hamiltonian in Eq.~(1) given
in the main article. When there is no driving, $A=0$, the coupling
is $q=0$; thus the two levels are uncoupled and there are no
microwave-induced peaks. Moreover, when $t_{\mathrm{so}}=0$ one of
the coefficients $c_{i}$ is zero; thus $q=0$ and the driving field
cannot couple the two levels. Finally, the parameters in this work
satisfy the regime $J_{1}(x)>J_{n}(x)$ with $n>1$, consequently at
a given magnetic field the single-photon peak is stronger than the
$n$-photon peak. This observation is consistent with the
experimental data.

\subsection{Rate equations}

In the spin blockade regime the electrical transport takes place
through the charge-cycle
$(0,1)\rightarrow(1,1)\rightarrow(0,2)\rightarrow(0,1)$, where
($k$, $m$) refers to a state with $k$ ($m$) holes on dot 1 (dot
2). We consider the single-spin states
$c^{\dagger}_{2\uparrow}|0\rangle$,
$c^{\dagger}_{2\downarrow}|0\rangle$, as well as the two-hole
states that form the anti-crossing $|u_{1}\rangle$,
$|u_{2}\rangle$, and determine the density matrix $\rho(t)$ of the
DQD in the transformed frame (`rotating' frame). Following a
standard open-system approach~\cite{blum} the equation of motion
of $\rho(t)$ can be written in the form
\begin{equation}\label{masterblumm}
\frac{d\rho(t)}{dt} = -\frac{i}{\hbar} [h_{\mathrm{DQD}}, \rho(t)]
+ \mathcal{L}\rho(t),
\end{equation}
where the incoherent term $\mathcal{L}\rho(t)$ accounts for the
interaction of the DQD with the two leads which is treated to
second order in the dot-lead tunnel coupling (sequential
tunneling). In this approximation the transition rates between the
DQD eigenstates due to the coupling of the DQD with the leads
acquire a simple form~\cite{Giavaras}. The effect of the
transformation $U(t)$ on the DQD-lead interaction is ignored and
Eq.~(\ref{masterblumm}) can be solved analytically in the steady
state, e.g., when $d\rho(t)/dt=0$. In this effective model the
electrical current through the DQD is proportional to the
population of the $|S_{02}\rangle$ state, which is extracted
directly from the populations of $|u_{1}\rangle$ and
$|u_{2}\rangle$.

\section{Floquet model}

The effective two-level model described in the preceding section
takes into account only the states which form the anti-crossing
point and neglects the time-dependent oscillating terms in the
Hamiltonian. In the spin blockade charge-cycle all triplet states
are relevant~\cite{Giavaras}, and in the limit $B\rightarrow 0$
the triplet states become quasi degenerate, thus the effective
model is questionable. Therefore, to test the overall accuracy of
the effective model, we describe in this section another model
that takes into account all the states which are involved in the
transport through the DQD~\cite{note1}, and treats the time
dependence of the DQD Hamiltonian $H_{\mathrm{DQD}}(t)$ exactly
within the Floquet formalism~\cite{platero, chu, kohler}.


\subsection{Floquet Hamiltonian}

The Hamiltonian of the DQD is periodic
$H_{\mathrm{DQD}}(t)=H_{\mathrm{DQD}}(t+T)$, with $T=2\pi/\omega$
being the period of the external electric field. For this reason
it is convenient to apply the Floquet formalism which is a
powerful tool for time-dependent periodic systems~\cite{platero,
chu, kohler}. According to the Floquet theorem, a solution of the
time-dependent Schr\"odinger equation with a periodic Hamiltonian
can be written in the form
\begin{equation}\label{timestat}
|\psi_j(t)\rangle = \exp\left( - i \frac{\epsilon_j t}{\hbar}
\right) |\phi_j(t)\rangle,
\end{equation}
where $|\phi_j(t)\rangle$ are the Floquet modes which have the
periodicity of the Hamiltonian, i.e.,
$|\phi_j(t)\rangle=|\phi_j(t+T)\rangle$, and $\epsilon_j$ are the
Floquet energies. These are time independent and can be defined,
for instance, within the interval $-\hbar\omega/2 < \epsilon_j <
+\hbar\omega/2$. The Floquet modes and energies satisfy the
following eigenvalue problem~\cite{brune}
\begin{equation}\label{floq}
\left( H_{\mathrm{DQD}}(t) - i \hbar \frac{\partial}{\partial t}
\right) |\phi_{j}(t)\rangle = \epsilon_{j} |\phi_{j}(t)\rangle,
\end{equation}
that is solved by expanding the time periodic
$H_{\mathrm{DQD}}(t)$ and $|\phi_{j}(t)\rangle$ in a Fourier
series:
\begin{eqnarray}
[H_{\mathrm{DQD}}(t)]_{nm} & = & \sum_{k}e^{ik\omega t}
[H^{k}_{\mathrm{DQD}}]_{nm}, \\
|\phi_{j}(t)\rangle & = & \sum_{k}
e^{ik\omega t} |\phi^{k}_{j}\rangle.
\end{eqnarray}
If we denote by $|y_{i}\rangle$ the basis vectors spanning the DQD
Hilbert space, and expand $|\phi^{k}_{j}\rangle$ in that basis
\begin{equation}
|\phi^{k}_{j}\rangle = \sum^{\mathcal{N}}_{i=1} W^{k}_{i,j}
|y_{i}\rangle,
\end{equation}
the eigenvalue problem Eq.~(\ref{floq}) becomes
\begin{equation}
\sum^{\mathcal{N}}_{l=1} \sum_{k} \left(  [ H^{n-k}_{\mathrm{DQD}}
]_{il} + n \hbar \omega \delta_{nk} \delta_{il} \right)
W^{k}_{l,j}  = \epsilon_{j} W^{n}_{i,j}.
\end{equation}
Here the indexes $n$, $k$ refer to the Fourier series, and the
indexes $i$, $l$ refer to the basis vectors. For the numerical
computations, this infinite system of coupled equations is
truncated to a finite but sufficiently large value to ensure good
convergence of the results.

\subsection{Master equation}

In the Floquet formalism, the density matrix $\rho(t)$ of the DQD
is expressed in the time-dependent Floquet basis
$|\phi_{j}(t)\rangle$, simplifying drastically the calculation of
the steady state~\cite{platero, chu, kohler}. Within the Born and
Markov approximations, the matrix elements $\rho_{ij}(t)$ satisfy
the master equation
\begin{eqnarray}
& & \hspace*{-0.5in} -\left(  \frac{\partial}{\partial t} + \frac{i}{\hbar}
\epsilon_{ij}\right)\rho_{ij}(t) = \sum_{k l} \rho_{lj}(t)
X_{ik;lk}(t) \nonumber \\
& & + \sum_{k l} \rho_{ik}(t) G_{lj;lk}(t) \nonumber \\
& & - \sum_{k l} \rho_{kl}(t)[
G_{ik;jl}(t) + X_{lj;ki}(t)]. \label{Fmaster}
\end{eqnarray}
with $\epsilon_{ij} = \epsilon_{i} - \epsilon_{j}$, and the
transition rates $X(t)$ and $G(t)$ quantify the interaction of the
DQD with the two leads. For simplicity, here we focus only on
$X(t)$ and consider only the interaction of dot 1 with the left
lead. The coupling of the DQD to the right lead can be treated in
a similar manner. The rate $X_{ij;kl}(t)$ is defined by the
Fourier expansion
\begin{equation}\label{rate}
X_{ij;kl}(t) = \sum_{K} e^{iK\omega t} X_{ij;kl}(K),
\end{equation}
\begin{equation}
\begin{split}
X_{ij;kl}(K)=&\Gamma\sum_{M \sigma}[c_{1\sigma}(K+M)]_{ij}
[c_{1\sigma}(M)]^*_{kl}f_{L}(\epsilon_{lk}-M\hbar\omega)\\
+&\Gamma\sum_{M \sigma}[c_{1\sigma}(-K-M)]^*_{ji}
[c_{1\sigma}(-M)]_{lk}f^{-}_{L}(\epsilon_{kl}+M\hbar\omega),
\end{split}
\end{equation}
where $f_{L}$ is the Fermi distribution at the chemical potential
of the left lead and $f^{-}_{L}=1-f_{L}$. The subband index is
defined by the index $M$. The DQD-lead coupling constant $\Gamma$
is proportional to $t^{2}_{L}$, and the matrix element is defined
through its Fourier transform as follows
\begin{equation}
[c_{1\sigma}(M)]_{ij}=\frac{1}{T}\int^{T}_{0} e^{-i M \omega t}
\langle\phi_{i}(t)|c_{1\sigma}|\phi_{j}(t)\rangle dt.
\end{equation}
For any two system operators $s_p$ and $s_w$, with
$s^{\dagger}_{p}= s_{w}$, the corresponding matrix elements
satisfy $[s_p(-M)]^{*}_{ji} = [s_w(M)]_{ij}$. To solve
Eq.~(\ref{Fmaster}) we assume that in the long-time limit the
density matrix, which describes the steady state, has the same
periodicity as the DQD Hamiltonian, thus it can be expressed in
the form
\begin{equation}\label{rsteady}
\rho_{ij}(t) = \sum_{N} e^{iN\omega t} \rho_{ij}(N).
\end{equation}
Substituting Eq.~(\ref{rate}) and Eq.~(\ref{rsteady}) into
Eq.~(\ref{Fmaster}) results in an infinite set of coupled
equations that is solved numerically by truncating $N$ to a finite
value. Having determined the steady state, the tunneling current
is computed by taking the average of the current operator $I = e i
[H, N_{R}]/\hbar$, where $N_{R} =
\sum_{k\sigma}d^{\dagger}_{Rk\sigma} d_{Rk\sigma}$ is the number
of holes in the right lead.

Figure~2 shows the background current and the microwave-induced
peak height near the $T_{+}$--$S$ anti-crossing point for a
microwave amplitude $A=30$ $\mu$eV. The basic features are in good
overall agreement with the experimental data [see main article
Fig.~2(c) and (d)]. The height of the current peaks is sensitive
to the DQD-lead coupling $\Gamma$ and the microwave amplitude $A$.
When $\Gamma$ is strong, $A$ has to be large for the peaks to be
visible. However, the computational time increases quickly with
$A$, because the Fourier expansions need extra terms to converge.
For this reason, to keep the numerical problem tractable we choose
$\Gamma$ in the GHz range.

A more detailed fit to the background current can be achieved by
coupling the DQD to a bosonic bath and introducing spin
flips~\cite{chorley11, giavaras20}. This approach offers limited
additional insight into the present experimental data, whilst
extra parameters have to be introduced to specify the spectral
density of the bath. Therefore, this approach is not pursued in
this work. Three-body states which for simplicity are not
accounted for in our model can also have some contribution to the
background current~\cite{giavaras20}.

Finally, we mention that the Floquet model can also be used to
assess the rotating wave approximation [Eqs.~(\ref{qq1}),
(\ref{qq2})] in the effective two-level model. In this case the
Hamiltonian $H_{\mathrm{DQD}}(t)$ in Eq.~(\ref{floq}) has to be
replaced by $h^{'}_{\mathrm{DQD}}(t)$ [Eq.~(\ref{2by2})]. The two
models are in agreement.

\begin{figure}[t]
\begin{center}
\includegraphics[width=0.45 \textwidth, keepaspectratio]{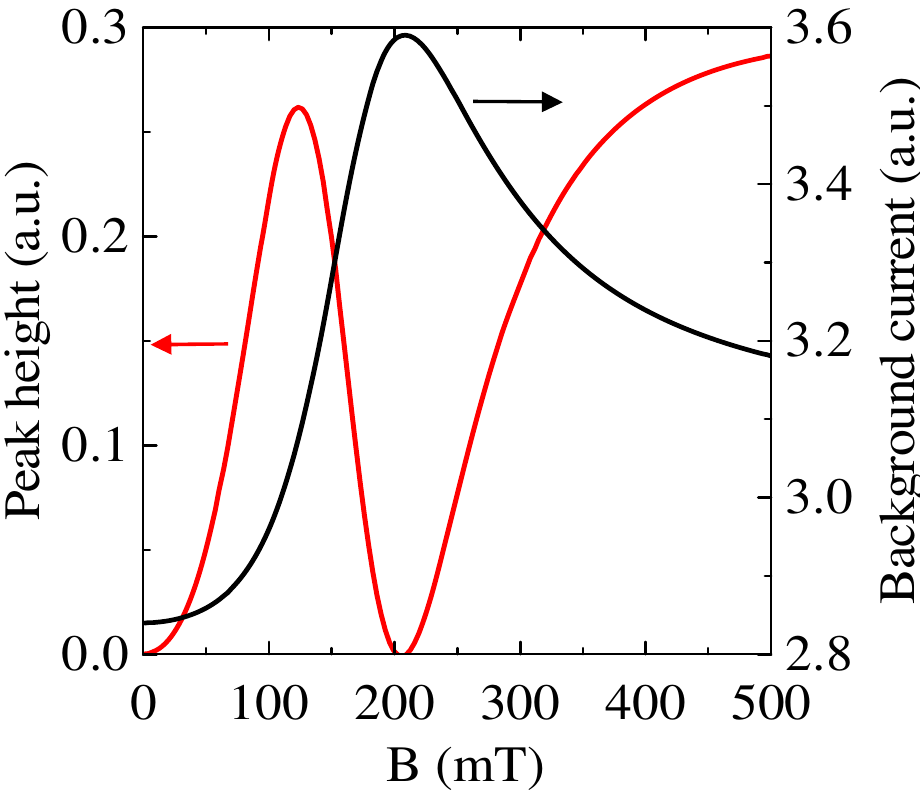}
\caption{Peak height (bright line, left axis) for a microwave
amplitude $A=30$ $\mu$eV, and background current without the
microwave field (dark line, right axis) as a function of the
magnetic field near the $T_{+}$--$S$ anti-crossing point. The
results are derived using the Floquet model described in the
Supplement. See also Figs.~2(c) and (d) in the main article.}
\end{center}
\end{figure}

\section{Double quantum dot parameters}

The experimental data suggests that the charging energies of the
two dots are $U_1\approx25$ meV and $U_2\approx5$ meV and the
$g$-factor is $g\approx1.8$ (see main article). In the
calculations we take for the two dots $g_{1}=g_{2}$, though this
assumption is not important. In the experiment an anti-crossing
point is probed at about $200$ mT and the anti-crossing gap is
about 1 GHz, but the exact values of the parameters $\delta$,
$t_{\mathrm{c}}$, and $t_{\mathrm{so}}$ are unknown. Consequently,
for the calculations we choose $\delta$, $t_{\mathrm{c}}$, and
$t_{\mathrm{so}}$ in order to form an anti-crossing point as in
the experiment, and simultaneously to achieve a good qualitative
agreement between the calculated and the measured background
currents ($A=0$). The SOI Hamiltonian $H_{\mathrm{so}}$ forms two
anti-crossing points, but the observed spectra indicate that only
one point is relevant for the chosen ranges of the magnetic field
and the driving frequency. The choice of the parameters $\delta$,
$t_{\mathrm{c}}$, and $t_{\mathrm{so}}$ is not unique and we
choose different values in the two models in order to achieve a
good fit to the background current. In the two-level model, the
parameters are $\delta=-1.85$ meV, $t_{\mathrm{c}}=0.135$ meV, and
$t_{\mathrm{so}}=0.15t_{\mathrm{c}}$; and in the Floquet model the
parameters are $\delta=-1.98$ meV, $t_{\mathrm{c}}=0.14$ meV, and
$t_{\mathrm{so}}=0.14t_{\mathrm{c}}$. Here, we present results for
$\delta<0$, but the models can also produce the general
experimental features for $\delta>0$.


\newpage 


\begin{thebibliography}{}
\bibitem{note1}
We neglect three-body states and consider in total 11 many-body
states.
\end{thebibliography}
\end{document}